\documentclass[twocolumn,preprintnumbers,amsmath,amssymb,superscriptaddress]{revtex4}
\usepackage{graphicx}
\usepackage{dcolumn}
\usepackage{bm}

\begin{document}

\newcommand{\be}{\begin{equation}}
\newcommand{\ee}{\end{equation}}
\newcommand{\ber}{\begin{eqnarray}}
\newcommand{\eer}{\end{eqnarray}}
\newcommand{\tr}{{\rm tr}}

\title{Statistics of lines of natural images and implications for visual detection}

\author{Ha Youn Lee} 
\affiliation{Department of Physics, Massachusetts Institute of Technology,
Cambridge, Massachusetts 02139}
\affiliation{Department of Physics, The Ohio State University, Columbus, Ohio
43210}
\author{Mehran Kardar}
\affiliation{Department of Physics, Massachusetts Institute of Technology,
Cambridge, Massachusetts 02139}

\begin{abstract} 
\indent 
As borders between different regions, lines are an important element of natural images.
Already at the level of the  mammalian primary visual cortex (V1), neurons respond best 
to lines of a given orientation.
We reduce a set of images to linear segments and analyze their statistical properties.
In particular, appropriately defined Fourier spectra show more power in their  {\em transverse} 
component than in the {\em longitudinal} one.
We then characterize filters that are best suited for extracting information
from such  images, and find some qualitative consistency with neural connections in V1. 
We also demonstrate that such filters are efficient in reconstructing missing lines
in an image.
\end{abstract} 

\maketitle

\indent 
An image on a  screen is represented by a set of intensities at each pixel.
The photoreceptors of the retina also respond to the intensity of light arriving
from specific directions.
However, when it comes to interpreting the content of an image, primary
clues are the borderlines between different regions.
Indeed, already at the level of the mammalian primary visual cortex (V1), neurons respond
best not to points of light, but to lines of particular orientation\cite{Hubel}.
It is thus important to inquire about the statistics of lines in natural scenes, and implications for vision.
In Ref.~\cite{Sigman}, such a study is performed
by first converting images to a set of lines:
Correlations of a pair of such lines with their relative location in space,
indicates a tendency towards {\em co-circularity},
namely the  most likely arrangement of the two segments is to lie along a circular arc joining them.
We start with a similar decomposition of images to lines, examine their statistics
(e.g. by Fourier transformation), and explore their implications for visual detection.

There are previous studies of the power spectrum of the (scalar) intensity correlations
of natural images~\cite{Field87,Ruderman94},
which find indications of scale invariance. 
For a vectorial quantity, a natural decomposition is into
longitudinal/transverse Fourier components, which measure the variations
parallel/perpendicular to a wavevector $\vec k$.
Such decomposition is for example quite common in studies of turbulent velocity fields~\cite{Monin,Arad}.
We construct similar measures of variations of the lines in natural images
(which unlike a vector field do not point to a specific direction),
and find enhanced power in the orthogonal (transverse) channel.
We designate this feature, related to the prevalence of sharp lines,
the  `transversality' of natural images.

Since the task of the visual cortex is to decipher visual signals,
its design is likely to depend upon statistics of natural images.
The visual input from the retina is carried by the optic nerve to
the lateral geniculate nucleus (LGN),
and then transferred to V1.
A prominent feature of neurons in V1 is that they respond most strongly when
viewing lines of a specific slant.
This orientation preference (OP) is thought to arise from the arrangement of the
feed-forward connections from the LGN~\cite{Hubel}.
However, within V1 there are also {\em horizontal connections} (extending for 2-5 mm)
which mostly link columns of neurons with similar OPs~\cite{Gilbert89,Malach93}.
Staining experiments with injected biochemical tracers in the tree shrew 
reveal that these lateral connections
are longer  and stronger along an axis in the map of visual field that corresponds
to the preferred orientation of the injection site~\cite{Bosking97}.
Similarly, in the cat visual cortex, facilitatory effects occur only between neurons which 
are co-axial in the spatial domain and co-oriented in the orientation domain~\cite{Nelson85}.

Although less understood than the feed-forward connections from LGN,
the long range connections in V1 are presumed to mediate the global integration of an image from its local elements.
Evidence supporting this comes from fMRI  investigations in monkeys and humans:
The neurons in V1 show higher response when  viewing a long extended line,
compared to randomly oriented segments of the line~\cite{Kourtzi}.
Here, we address the characteristics of the lateral connections
from the perspective of information theory~\cite{Laughlin,Atick1990,Atick1992,Dan,Bialek,Kardar}.
Using the two point correlation functions for lines
in natural images, we construct long-range filters that are
optimally suited for harvesting visual information.
We find that the strongest connections are between neurons with a
common OP directed along the line joining the visual field locations of the neurons,
as observed in the cortex of cat and tree shrew.  

The long-range connections that maximize information reinforce the local (feed-forward)
input to a neuron.
If the local signal is for any reason corrupted, the global information can help to reconstruct it.
Indeed psycho-physical tests show the facility of the brain to
recognize missing segments in an image~\cite{Grossberg}.
To mimic this effect,   we construct filters that are optimally
suited to study images composed of {\em directed} lines.
Since most of the information is in the `transverse' channel, these filters have
a transverse character.
We demonstrate that transverse filters perform much better than isotropic ones in
reconstructing missing gaps in simple images.

To obtain statistics of lines, we start with
a collection of black and white pictures from a database, 
``http://hlab.phys.rug.nl/imlib/index.html,''\cite{Hans van Hateren}
which includes trees, buildings, flowers, leaves, and grass.
The data, which is in the form of a scalar intensity at each pixel, is then converted
into oriented  segments $[s_x(\vec{X}),s_y(\vec{X})]$ at each pixel $\vec{X}$,
using filters based on the second derivative of a Gaussian and its Hilbert transform\cite{Freeman}.
Since $[s_x,s_y]$ and $[-s_x,-s_y]$ describe the same orientation, 
we introduce the tensor field 
${\bf s}_{\alpha\beta}(\vec{X})=s_{\alpha}(\vec{X}) s_{\beta}(\vec{X})$, 
which is invariant under reflection.
The two dimensional Fourier transforms of the components of this tensor
lead to a corresponding ${\bf S}_{\alpha\beta}(\vec{k})$.
The longitudinal and transverse components of the tensor are then obtained as
\begin{equation}\label{eq: LTofS}
S_\ell(\vec k)=\tr\left[{\bf L}(\vec k){\bf S}(\vec k)\right],\qquad 
S_t(\vec k)=\tr\left[{\bf T}(\vec k){\bf S}(\vec k)\right],
\end{equation}
with the aid of the projection operators
\begin{equation}\label{eq:LTdefine}
{\bf L}_{\alpha\beta}(\vec k)=\hat{k}_{\alpha}~ \hat{k}_{\beta},\quad
{\bf T}_{\alpha\beta}(\vec k)=[\delta_{\alpha \beta}-\hat{k}_{\alpha}~ \hat{k}_{\beta}],
\end{equation}
where $\hat{k}$  is the {\em unit vector} in the direction of $\vec{k}$.

\begin{figure}[h]
\centering
\includegraphics[width=3.5cm,height=3.5cm]{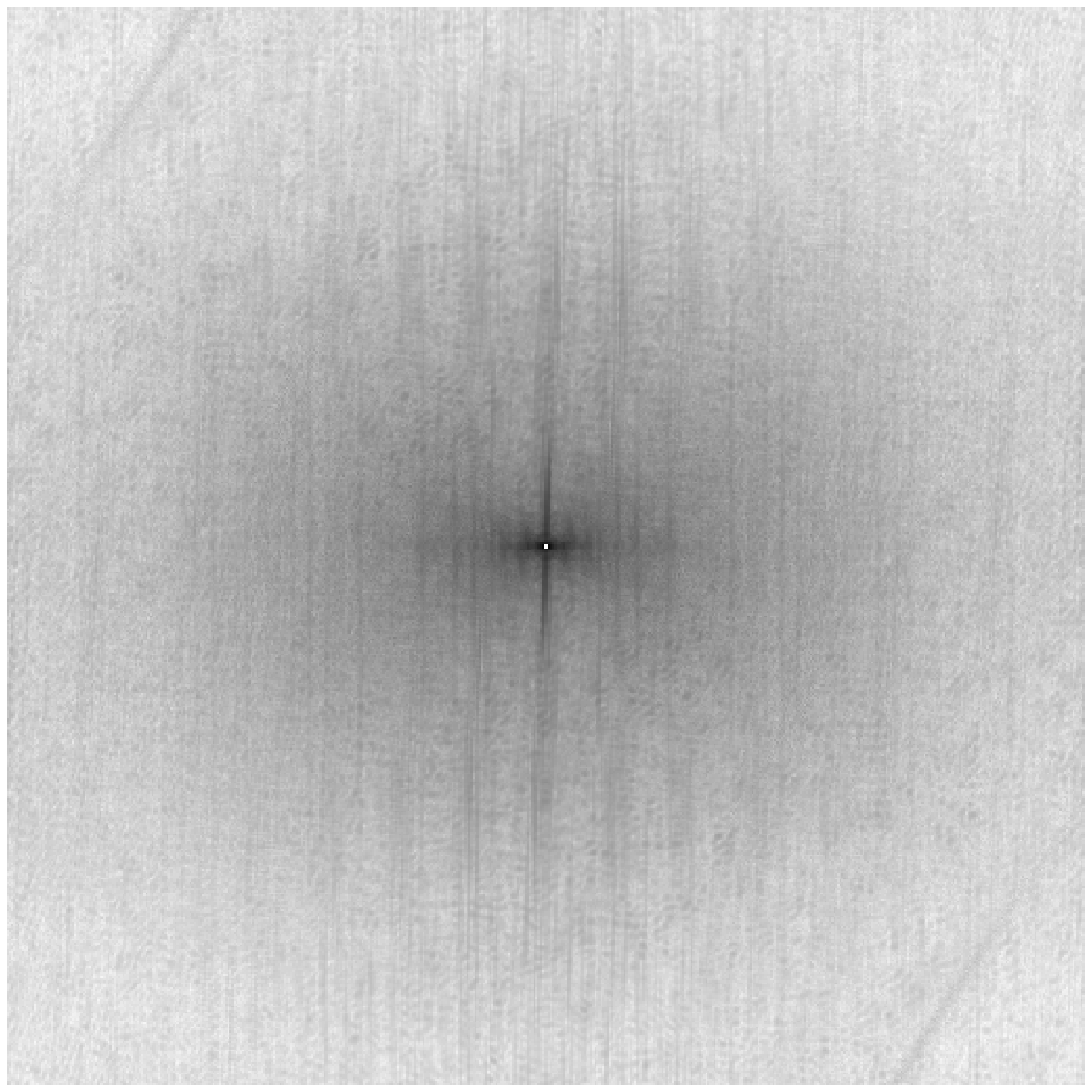}
\includegraphics[width=3.5cm,height=3.5cm]{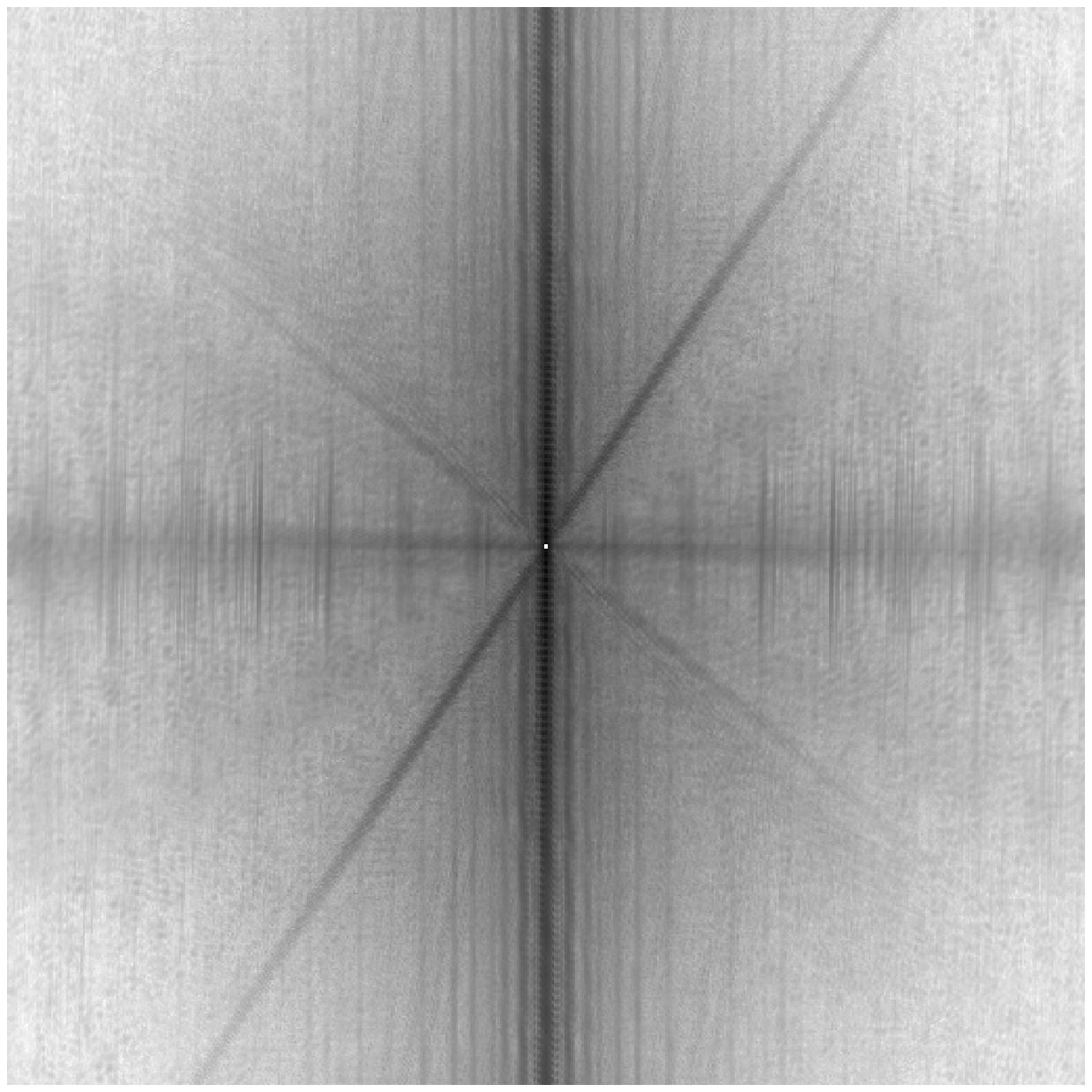}\\
(a)\hspace*{3.2cm}(b)\\
\includegraphics[width=6.5cm,height=4.5cm]{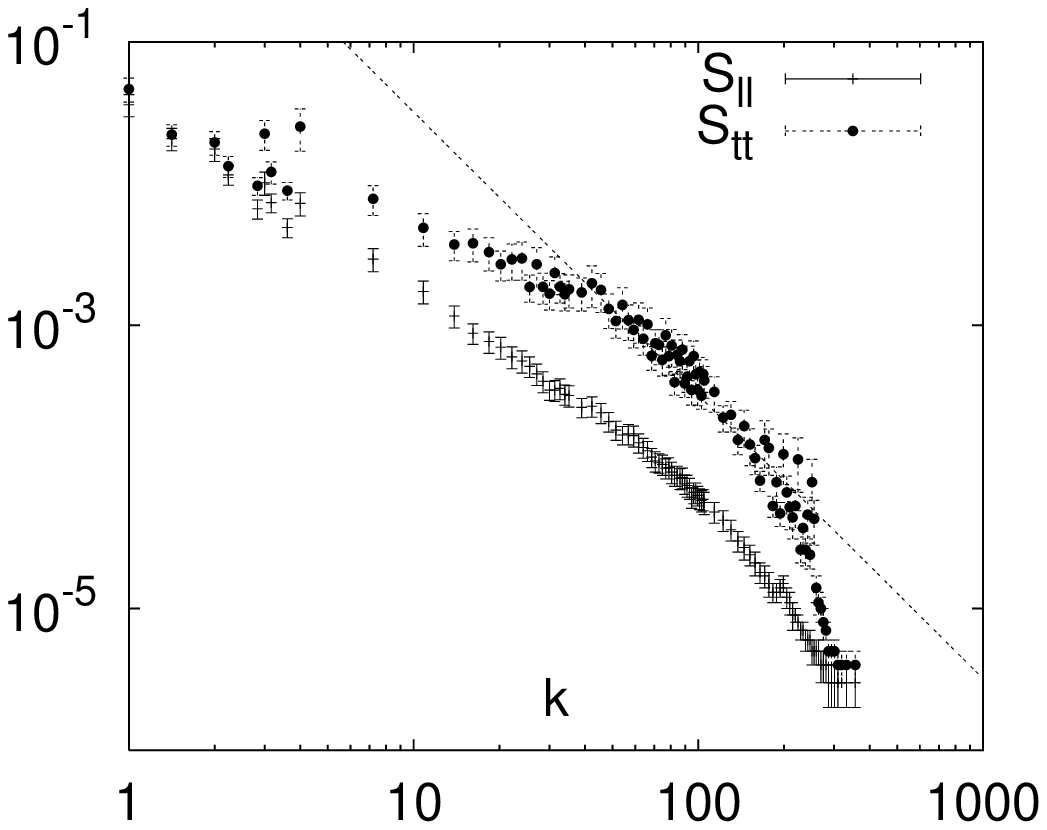}\\
(c)
\caption[]{
Intensity plots of the longitudinal $S_{\ell\ell}(\vec{k})$ (a), 
and transverse $S_{tt}(\vec{k})$ (b),
power spectra obtained from averaging over a set of 100 natural images.
(c) Log-log plots of $S_{\ell\ell}({k})$ and  $S_{tt}({k})$
after averaging over all angles.
}
\label{fig:out}
\end{figure}

Figures~\ref{fig:out} (a) and (b) show the power spectra
$S_{\ell\ell}(\vec{k})\equiv|S_\ell(\vec k)|^2$ and $S_{tt}(\vec{k})\equiv |S_t(\vec k)|^2$
after averaging over 100 images.
Clearly these quantities are not isotropic and vary with angle.
This is due to the predominance of vertical and horizontal segments in the images.
The bias of oriented segments along cardinal directions in natural scenes
is well known~\cite{Switkes}, and a similar bias exists in the OPs
of cortical maps from adult ferret and cat~\cite{Pettigrew,Chapman}.
There is a corresponding larger area of V1 devoted to vertical and horizontal orientations,
and a greater stability of cardinal neurons to changes of orientation~\cite{Dragoi}.
Since we are not interested in the predominance of specific orientations,
we remove this anisotropy by averaging over rotated images~\cite{rotated-image}.
Equivalently, we can average the power spectra in Fig.~\ref{fig:out} over all angles,
resulting in $S_{\ell\ell}$ and $S_{tt}$ as a function of $|\vec{k}|$,
as depicted in Fig.~\ref{fig:out}(c).

The data in Fig.~\ref{fig:out} clearly shows higher power in the transverse component.
As with the  {\em intensities} of natural images~\cite{Ruderman94},
the power spectra are reasonably close to a power-law form $1/k^{2-\eta}$,
presumably reflecting an underlying scale invariance since objects can appear
at any distance from the viewer.
(The straight line in Figs.\ref{fig:out}(c) corresponds to $\eta=0$.)
We believe that the deviations from scale invariance (especially pronounced in the
transverse component) are an artifact of our images.
Converting intensity data to orientations involves filters with an
inherent short distance scale;
at such short scales the two power spectra coincide as required by local isotropy.
There is a range of intermediate scales in which both spectra can
be fitted to power laws.
The deviations from scale invariance at shortest wavelengths are  
likely due to a tendency to frame photographs to include whole objects, 
excluding images with parts of objects extending beyond the frame~\cite{global}.

The enhanced transverse power is a consequence of the abundance of
sharp and extended edges in natural images.
An elementary illustration is obtained by comparing a long straight line
with a horizontal arrangement of short vertical segments as in a fence. 
The former has no longitudinal Fourier component
while the latter has weak transverse character.
Searching for other sets of images with different statistics, 
we did a sampling of paintings from modern art.
We find that many paintings from the impressionist school
with blurred
lines have approximately equal transverse and longitudinal powers.
By contrast,  cubist paintings with sharp lines
share (and in fact exceed) the transversality of natural images~\cite{WebData}.

We next attempt to relate the above statistics 
to the lateral connections between neurons of V1,
using information theoretic methods.
The general idea is to construct an {\em output signal} by removing redundant
correlations of the {\em input signals} as much as possible, maximizing the entropy of the output.
Information  theory has been used to describe early visual processing,
such as the contrast response of large monopolar cells\cite{Laughlin},
the `center surrounded' receptive fields in the retina\cite{Atick1990,Atick1992},
and the {\it white} spatial/temporal power spectrum of signals from the LGN\cite{Atick1992,Dan}.
In Ref.~\cite{Bialek}, filters for processing intensity  inputs to V1
were calculated by maximizing information subject to certain costs.
Our approach is based on the latter, and as extended in Ref.~\cite{Kardar},
but employing an input signal which is an orientation field.

The response of simple cells in V1 is primarily to an oriented line in a preferred direction,
which we shall approximate by $\tr[{\bf t}(\vec{x}){\bf s}(\vec{X})]=[\vec{t}(\vec{x})\cdot \vec{s}(\vec{X})]^2$.
Here ${\bf s}_{\alpha\beta}(\vec{X})=s_{\alpha}(\vec{X}) s_{\beta}(\vec{X})$ is
constructed from the orientation of the image segment (input signal) 
at position $\vec{X}$ in the visual field,
while a tensor ${\bf t}_{\alpha\beta}(\vec{x})\equiv t_{\alpha}(\vec{x}) t_{\beta}(\vec{x})$
is defined in terms of the OP of a neuron at location $\vec{x}$ in V1.
The topographic map between the visual field and V1 provides a mapping between $\vec{x}$
and $\vec{X}$. However, to emphasize that this mapping is not one to one, with many V1 neurons
responding to signals at the same position in the visual field, we use two symbols
$\vec{X}$ and $\vec{x}$.

Our main interest is in the {\em lateral connections} to a cell from other neurons in V1.
With this aim, we indicate the net response (neuron firing rate), by
\begin{equation}
O(\vec{x})=\tr[{\bf t}(\vec{x}) {\bf s}(\vec{X})]
+\int d^2 y F (\vec{x},\vec{y})\tr[{\bf t}(\vec{y}) {\bf s}(\vec{Y})]+ \eta(\vec{x}).
\label{eq:neuron-output}
\end{equation}
The `filter function' $F (\vec{x},\vec{y})$ denotes the strength of the horizontal
connection between the neurons at $\vec{x}$ and $\vec{y}$;
$\eta(\vec{x})$ is the noise experienced by the neuron which is assumed to be
uncorrelated at different points, with 
$\langle \eta(\vec{x}) \eta(\vec{x'})\rangle=\sigma^2 {\delta}^2 (\vec{x}-\vec{x'})$.
Given the stochastic nature of the input signal (as well as the noise), the output $O(\vec{x})$
is a random variable with a (joint) probability distribution $p[O(\vec{x})]$.
The associated Shannon information is
\begin{equation}
I=-\langle \ln p[O(\vec{x})]\rangle\approx  {1 \over 2} ~{\mbox {ln}} ~{\mbox {det}}
[\langle O(\vec{x})O(\vec{x'})\rangle_c],
\label{eq:inten-information}
\end{equation}
where $\langle O(\vec{x})O(\vec{x'})\rangle_c$ is the second cumulant (co-variance) of the output.
The final approximation assumes a Gaussian $p[O(\vec{x})]$, and ignores higher order cumulants.
For low signal to noise ratio we can further simplify the result to 
\begin{eqnarray}
I&\approx&{1 \over 2} \int d^2 \vec{x}~ {\cal S}_{\alpha\beta \gamma \delta}(0)~{\bf t}_{\alpha\beta}(\vec{x})
{\bf t}_{\gamma \delta}(\vec{x})  \\
&+&\hspace*{-0.3cm}\int d^2 \vec{x} \int d^2 \vec{y}~F(\vec{x},\vec{y})
{\cal S}_{\alpha\beta \gamma \delta}(\vec{X}-\vec{Y}) {\bf t}_{\alpha\beta}(\vec{x})
{\bf t}_{\gamma \delta}(\vec{y}), \nonumber
\label{eq:inten-approxinfo}
\end{eqnarray}
where 
${\cal S}_{\alpha\beta\gamma\delta}(\vec{X}-\vec{Y})=\langle {\bf s}_{\alpha\beta}(\vec{X}){\bf s}_{\gamma\delta}
(\vec{Y})\rangle_c/\sigma^2$ denotes the co-variance of the input signal.

To provide a meaningful comparison of different filters, we need
to maximize the above information subject to costs and constraints.
In particular, it is reasonable to assume that an expansion of the wiring costs for
small $F$ starts at quadratic order (so that no connections are formed in the absence
of any gain). Following Refs.~\cite{Bialek,Kardar}, we introduce a cost function
\begin{equation}
C[{\bf t}, F]=
C_1[{\bf t}]+
{1 \over 2} \int d^2 \vec{x} d^2 \vec{y}~C_2(\vec{x}-\vec{y}) F(\vec{x},\vec{y})^2,
\label{eq:cost}
\end{equation}
where $C_2(r)$ is a cost for connecting neurons at a separation $r$.
We would like to maximize $I-C$ with respect to both ${\bf t}(\vec{x})$ and $F(\vec{x},\vec{y})$.
The largest contribution should come from the local OPs encoded
in  ${\bf t}(\vec{x})$.
However, this is not our concern here, and for this reason we have not dwelled on the
precise form of   $C_1[{\bf t}]$.
Given that the field ${\bf t}_{\alpha\beta}(\vec{x})$ has somehow been established, we would like
to determine $F(\vec{x},\vec{y})$.
Assuming that the latter connections provide a small
correction to the overall information, maximization gives
\begin{equation}
F(\vec{x},\vec{y})= {{{\bf t}_{\alpha\beta}(\vec{x}) {\cal S}_{\alpha
\beta\gamma\delta}(\vec{X}-\vec{Y}) {\bf t}_{\gamma\delta}(\vec{y})} \over
C_2(\vec{x}-\vec{y})}. 
\label{eq:inten-filter}
\end{equation}

\begin{figure}[h]
\centering
\includegraphics[width=9cm,height=6cm]{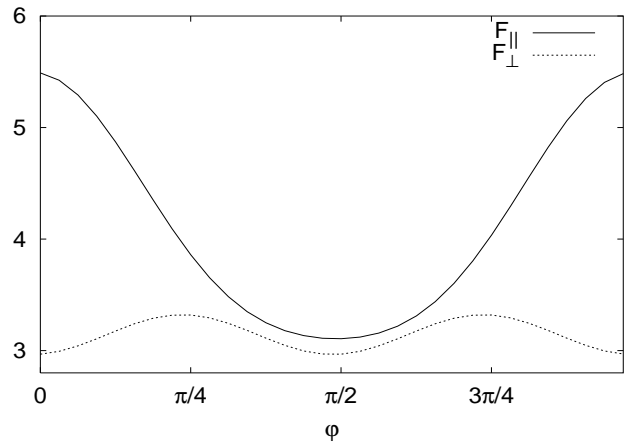}
\caption[]{
The strength of horizontal connections among neurons 
with parallel OPs (solid line $F_{||}$), and with orthogonal OPs (dotted line $F_\perp$),
as a function of their angle $\varphi$ to the line between their locations in the visual field.
The results are for a fixed separation, and obtained from the statistics of lines in a set
of five images of trees.
}
\label{fig:signal-correlation}
\end{figure}

The optimal connection between two V1 neurons thus
depends on their OPs, and the correlations in natural signals
at the corresponding locations and orientations.
This qualitatively agrees with the
observations in tree shrew~\cite{Bosking97} and cat~\cite{Nelson85}. 
To confirm that Eq.~(\ref{eq:inten-filter}) does indeed predict the
enhanced horizontal connections between colinear and co-oriented neurons, 
we measured the two point correlation functions
by averaging over a set of five images of trees.
Figure \ref{fig:signal-correlation} compares the strength of 
the connection among neurons with parallel OPs ($F_\parallel$)
to that of neurons with orthogonal OPs ($F_\perp$),
as a function of the angle $\varphi$ between one of the OPs,
and the line joining their locations in the visual field.
The figure is for a constant separation $|\vec x-\vec y|$;
the angular dependence is not very sensitive to this separation.
There is a strong maximum in $F_\parallel$ at colinearity $\varphi=0$;
while $F_\perp$ (which is always smaller than $F_\parallel$) shows weak
maxima at $\pi/4$ and $3\pi/4$ (consistent with the co-circularity principle\cite{Sigman}).

One advantage of optimal filters 
is observed by noting that
the resulting noise-average output of a neuron is
\begin{equation}
O(\vec{x})={\bf t}_{\alpha\beta}(\vec{x})\left[{\bf s}_{\alpha\beta}(\vec{X})+
\int d^2 y \frac{\langle {\bf s}_{\alpha\beta}(\vec{X}){\bf s}_{\gamma\delta}(\vec{Y})\rangle_c}{C_2(\vec{x}-\vec{y})}{\bf s}_{\gamma\delta}(\vec{y})\right].
\label{eq:optimal-output}\nonumber
\end{equation}
If the primary signal $s_\alpha(\vec{X})$ is somehow corrupted, the connections
provide a guess based on global statistics.
Let us employ similar principles to construct artificial algorithms for visual
detection, which (like the human brain) are adept at deducing global shapes in an image
composed of lines.
As an alternative to Eq.~(\ref{eq:neuron-output}) which avoids introduction of an OP
field,  
we define a {\em vectorial output} whose components are
\begin{equation}
O_{\alpha}(\vec{x})=\int d^2 y {\bf F}_{\alpha \beta} (\vec{x}-\vec{y})
s_{\beta} (\vec{y}) + \eta_\alpha(\vec{x}).
\label{eq:output-vector}
\end{equation}
The filter  
is now a $2\times 2$ matrix.
As in Eq.~(\ref{eq: LTofS})
its Fourier transform can be projected into longitudinal/transverse parts as
\begin{equation}
{\bf F}_{\alpha \beta}(\vec{k})= {\bf L}_{\alpha\beta}(\vec k)F_{\ell}(\vec{k})
+{\bf T}_{\alpha\beta}(\vec k) F_{t}(\vec{k}).
\label{eq:LTofF}
\end{equation}

\begin{figure}[h]
\centering
\includegraphics[width=10.cm,height=5cm]{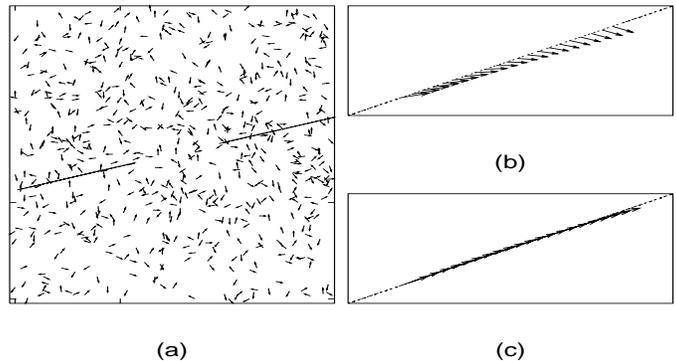}
\caption[]{
(a) A test image of a directed line with a gap (plus noise).
Reconstructions of the missing segment, with an {\em isotropic} filter (b);
and with a {\em transverse} filter (c).
}
\label{fig:anisofilter}
\end{figure}

Now consider a set of images in the form of a {\em vector field} $\vec{s}(\vec{x})$,
which is statistically invariant under translations.
For low signal to noise, the Shannon information in the output is
\begin{equation}
I={A \over 2 } \int {d^2 k \over {(2 \pi)}^2}
\left[|F_{\ell}(\vec{k})|^2 S_{\ell\ell} (\vec{k})
+|F_{t}(\vec{k})|^2 S_{tt}(\vec{k})\right] ,
\end{equation}
with projected signal correlations  defined as in Eq.~(\ref{eq:LTofF}).
As before, we can search for filters that maximize
information subject to specified costs.
However, to simplify matters we note that the transverse and longitudinal channels
can be treated independently, and that most of the information is in the
transverse channel which has the larger signal power spectrum.
As such, we compared the performance of the following filters:
{\bf (1)} A {\em transverse filter} with $F_{t}(\vec{k})=\phi(k)$ and $F_{\ell} (\vec{k})=0$;
and 
{\bf (2)} an {\em isotropic filter} with $F_{t}(\vec{k})=F_{\ell} (\vec{k})=\phi(k)/\sqrt{2}$.
In both cases, we chose  $\phi(k)\propto \exp(-k^2/4C)$.
The input image, illustrated in Fig.~\ref{fig:anisofilter}(a) consists of vectors,
some pointing randomly (noise), and some arranged into a line with a gap
 (corrupted image).
Figures~\ref{fig:anisofilter}(b)-(c) indicate how well the filters reconstruct 
the missing part.
The output of the transverse filter is both stronger and better oriented to the
erased line.
(Detailed results quantifying the improvements shall be reported elsewhere.)

The authors acknowledge support from the NSF grant  DMR-01-18213 (MK);
and by a  University Postdoctoral Fellowship
 from The Ohio State University (HYL).


\begin{thebibliography}{7}
\bibitem{Hubel} D.~H. Hubel and T.N. Wiesel, J. Physiol. {\bf 160}, 215 (1962).
\bibitem{Sigman} M. Sigman, G. A. Cecchi, C. D. Gilbert, and M. O. Magnasco, 
Proc. Natl. Acad. Sci. USA {\bf 98}, 1935 (2001).
\bibitem{Field87} D. J. Field, 
J. Opt. Soc. Am. A {\bf 4}, 2379 (1987).
\bibitem{Ruderman94} D. Ruderman and  W. Bialek, 
Phys. Rev. Lett. {\bf 73}, 814 (1994).
\bibitem{Monin}
A. S. Monin and A. M. Yaglom, {\it Statistical Mechanics}
(MIT, Cambridge, 1971), Vol. 2 pp. 1-58.
\bibitem{Arad}
I. Arad, {\em et. al},
Phys. Rev. Lett. {\bf 81}, 5330  
(1998).
\bibitem{Gilbert89} C. D. Gilbert and T. N. Wiesel,
J. Neurosci. {\bf 9}, 2432 (1989).
\bibitem{Malach93} R. Malach, Y. Amir, M. Harel, and A. Grinvald, 
Proc. Natl. Acad. Sci. USA {\bf 90}, 10469 (1993).
\bibitem{Bosking97} W. H. Bosking, Y. Zhang, B. Schofield, and D. Fitzpatrick, 
J. Neurosci. {\bf 17}, 2112 (1997).
\bibitem{Nelson85} J. I. Nelson, and B. J. Frost, 
Exp. Brain Res. {\bf 61}, 54 (1985).
\bibitem{Kourtzi} Z. Kourtzi, {\em et al.},
Neuron {\bf 37}, 333 (2003).
\bibitem{Laughlin} S. B. Laughlin, 
Z. Naturf.
{\bf 36c}, 910 (1981).
\bibitem{Atick1990} J. J. Atick and A. N. Redlich, 
Neural Comput.
{\bf 2}, 308 (1990).
\bibitem{Atick1992} J. J. Atick, 
Network: Comput. Neural Sys.
{\bf 3}, 213 (1992).
\bibitem{Dan} Y. Dan, J. J. Atick, and R. C. Reid, 
J. of Neurosci.
{\bf 16}, 3351 (1996).
\bibitem{Bialek} W. Bialek, D. L. Ruderman, and A. Zee, 
in {\it Advances in Neural Information Processing Systems,}
R.P. Lippman, editor (Morgan Kaufmann, San Mateo, CA 1991), p. 363.
\bibitem{Kardar} M. Kardar and A. Zee, 
Proc. Natl. Acad. Sci. USA {\bf 99}, 15894 (2002).
\bibitem{Grossberg} S. Grossberg and E. Mingolla,
Psychol. Rev. {\bf 92}, 173 (1985).
\bibitem{Hans van Hateren}
J. H. van Hateren and A. Van der Schaaf,
Proc. R. Soc. London B {\bf 265}, 359 (1998).
\bibitem{Freeman} W. T. Freeman and E. H. Adelson,
IEEE Trans. Patt. Anal. Mach. Intell. {\bf 13}, 891 (1991).
\bibitem{Switkes} E. Switkes, M. J. Mayer, and J. A. Sloan, 
Vision Res. {\bf 18}, 1393 (1978).
\bibitem{Pettigrew} J. D. Pettigrew, T. Nikara, and P. O. Bishop, 
Exp. Brain Res. {\bf 6}, 373 (1968).
\bibitem{Chapman} B. Chapman and T. Bonhoeffer, 
Proc. Natl. Acad. Sci.USA  {\bf 95}, 2609 (1998).
\bibitem{Dragoi} V. Dragoi, C. M. Turch, and M. Sur, 
Neuron {\bf 32}, 1181 (2001).
\bibitem{rotated-image}
We confirmed that
the spectra become more isotropic
as we  average over more rotated images.
Note that with a matrix ${\bf S}_{\alpha\beta}$ obtained from an orientation field,
there is no a priori reason for the cross correlations $S_{lt}(\vec{k})$ and $S_{tl}(\vec{k})$ 
to be zero. We do find that these correlations are small, and  also decrease
as we average over rotated images.
\bibitem{global} This was tested by generating random lines within one frame.
As the length of lines increases, the meeting point between the
two spectra is shifted to smaller $k$.
\bibitem{WebData}
Additional pictures and data are available online from
http://www.mit.edu/$\sim$kardar/research/transversality\\/ModernArt/.


\end{thebibliography}
\end{document}